\documentstyle[12pt]{article}

\title{Dynamics of an Acoustic Polaron in One-Dimensional Electron-Lattice 
System}

\author{
Yoshitaka {\sc Arikabe}, Makoto {\sc Kuwabara}$^1$ and Yoshiyuki {\sc Ono}
}

\begin{document} 

\sloppy
\maketitle

\medskip
\begin{center}
{\it 
Department of Physics, Toho University, \\
Miyama 2 - 2 - 1, Funabashi, Chiba 274 \\
$^1$Fundamental Physics Section, Physical Science Division, \\
Electrotechnical Laboratory, \\
Umezono 1 - 1 - 4, Tsukuba, Ibaraki 305
}
\end{center}

\medskip

\centerline{(received: January 9, 1996)}

\medskip

{\small 
The dynamical behavior of an acoustic polaron in typical non-degenerate 
conjugated polymer, polydiacetylene, is numerically studied by using 
Su-Schrieffer-Heeger's model for the one dimensional electron-lattice system. 
It is confirmed that the velocity of a polaron accelerated by a constant 
electric field shows a saturation to a velocity close to the sound velocity 
of the system, and that the width of a moving polaron decreases as a 
monotonic function of the velocity tending to zero at the saturation 
velocity. The effective mass of a polaron is estimated to be about one 
hundred times as heavy as the bare electron mass. Furthermore the linear 
mode analysis in the presence of a polaron is carried out, leading 
to the conclusion that there is only one localized mode, i.e. the 
translational mode. This is confirmed also from the phase shift of extended 
modes. There is no localized mode corresponding to the 
amplitude mode in the case of the soliton in polyacetylene. 
Nevertheless the width of a moving polaron shows small oscillations in 
time. This is found to be related to the lowest odd symmetry extended 
mode and to be due to the finite size effect. 
}

\newpage

\section{Introduction}

Polydiacetylene (PDA) is one of typical quasi-one-dimensional materials, 
which can exist as perfect single crystals with parallel straight chains 
of polymers separated by a rather long distance ($\sim$ 0.7 nm). About 
one and a half decades ago, Donovan and Wilson~\cite{rf:1} studied 
experimentally the dynamical properties of excess charges created by 
photo-excitation. What they found was unexpectedly interesting: the 
excess charges have very high mobility larger than 20m$^2$s$^{-1}$V$^{-1}$, 
yet their velocity is a constant of about half of the sound velocity 
for the field range 10$^2$--10$^6$ Vm$^{-1}$. Wilson~\cite{rf:2} later 
proposed the transport mechanism due to an acoustic polaron in order to 
explain these experimental facts. He discussed the dynamics of the 
acoustic polaron in terms of Su-Schrieffer-Heeger's (SSH) model.~\cite{rf:3} 
The acoustic polaron is created as the ground state when there is one 
electron in the conduction band. The electron is trapped by the local 
lattice distortion which is introduced by the presence of the electron 
itself. The electronic state and the lattice distortion are determined 
by solving a self-consistent equation. Wilson solved this self-consistent 
equation analytically within the continuum approximation for the SSH model. 
In solving the equation for a moving polaron, he assumed the space and time 
dependence of the lattice distortion and/or the electronic wave function 
could be described in terms of $x-vt$ with $v$ the velocity of the polaron. 
Wilson's solution indicates that the velocity shows a saturation at the 
sound velocity of the system, that the total energy diverges as a function of 
the velocity at the saturation velocity, and that the polaron width decreases 
with increasing velocity and vanishes at the saturation velocity. 
These results are consistent with the experimental ones by Donovan and Wilson. 

Recently Ono and co-workers~\cite{rf:4,rf:5,rf:6,rf:7} have developed a 
method of a numerical simulation to study the dynamics of a soliton in 
polyacetylene(PA). 
In this formulation, we prepare the state bearing a soliton by numerically 
calculating the lattice distortions and the electronic wave functions by 
solving a set of self-consistent equations derived from the condition of 
minimizing the total energy of the system and the soliton is accelerated by 
a physical force. Therefore we need not employ any assumption about the 
functional shape of a moving soliton. The time development of the lattice 
distortions and the electronic wave functions are obtained by solving the 
equation of motion and the time-dependent Schr\"odinger equation. 
Application of this method to the dynamics of an acoustic polaron is 
straightforward.  It will be worthwhile to study the dynamics of the acoustic 
polaron in the light of this method. 

We have learned from the studies of dynamical behaviors of solitons and 
polarons in PA that the linear mode analysis is quite useful to 
understand the dynamics of such nonlinear localized 
excitations.~\cite{rf:8,rf:9,rf:10,rf:11,rf:12,rf:13} 
In the present paper, we also carry out the linear mode analysis 
in the presence of an acoustic polaron and discuss the relation 
of the linear vibrational modes and the behaviors of a moving polaron.  
In the case of the solitons and polarons in PA, where they are 
related to the optical lattice vibrations due to the fact that the electrons 
are half-filling the $\pi$-band, the number of localized modes and the 
phase shift of the extended modes in the limit of vanishing wave number are 
related just as in Levinson's theorem for the one-dimensional Schr\"odinger 
equation.~\cite{rf:11,rf:12,rf:14} It will be discussed whether 
such an analogy is valid in the case of the acoustic polaron. 

In the following section, the model and the method of calculation are briefly 
described.  The results of simulations are presented in \S~3. In \S~4, 
the results of the linear mode analysis are explained and particularly 
the relation between the number of localized modes and the behavior of 
the phase shift of extended modes is discussed. The role of linear modes 
in the dynamics of the polaron is considered. We present concluding 
remarks and discussions in the final section.

\section{Model and Method of Calculation}

Although we are bearing in mind the non-degenerate conjugated polymer, 
PDA , we employ the SSH model which is widely used in studying 
the degenerate conjugated polymer PA. This will be 
allowed since we are treating a single carrier excited in the bottom 
region of the conduction band. PDA has a much more complicated 
chemical structure than PA, but such details will not matter in 
discussing the basic dynamics of an acoustic polaron. 

The SSH model Hamiltonian is expressed as follows,
\begin{eqnarray}
H_{\rm SSH} & = & H_{\rm ele} + H_{\rm lat}, \label{eq2.1} \\
H_{\rm ele} & = & -{\sum_{n}}
              (t_0-\alpha y_n)[{\rm e}^{{\rm i}\gamma A}c^\dagger_{n}
         c_{n+1}+{\rm e}^{-{\rm i}\gamma A}c^\dagger_{n+1}c_{n}], \nonumber \\
         & & \label{eq2.2} \\
H_{\rm lat} & = & {K \over 2}{\sum_n}y_n^2 + {M \over 2}{\sum_n}{\dot u}_n^2, 
\label{eq2.3}
\end{eqnarray}
where $y_n \equiv u_{n+1}-u_n$ and $\gamma = ea/\hbar c$ with $u_n$ the 
displacement of the $n$-th site, $a$ the lattice constant, $e$ the absolute 
value of the electronic charge and $c$ the light velocity. The field 
operators $c_n$ and $c_n^\dagger$ annihilate and create an electron on the 
$n$-th site; the spin index is omitted since we consider single electron 
problem. $K$ is the force constant mainly due to the $\sigma$-bond, $M$ the 
mass of a unit cell, $t_0$ the nearest neighbor transfer integral of a 
$\pi$-electron, and $\alpha$ the electron-lattice coupling constant.  
Here we have introduced similarly as in ref.~\cite{rf:4} a time 
dependent vector potential $A$ which is related to a uniform electric field 
$E$ as $E = -\dot{A}/c$.

The standard values of parameters accepted generally for PDA~\cite{rf:2} 
are $t_0 = 0.64$eV, $K= 2.4$eV/\AA$^2$, $\alpha = 0.38$eV/\AA and 
$a= 4.9$\AA.  In this paper we basically use these values as far as we 
do not mention other values. For these values of parameters, the bare 
optical phonon frequency $\omega_Q (= \sqrt{4K/M})$ is equal to 
1.47$\times 10^{13}$sec$^{-1}$ and the sound velocity 
$v_{\rm s}(=\omega_Qa/2)$ is 3.6$\times 10^5$cm/sec. In the following, 
the time is scaled by $\omega_Q^{-1}$, the velocity by $v_{\rm s}$ and 
the length by $a$. 
It is most convenient to scale the strength of the field by 
\begin{equation}
E_0 \equiv {{\hbar \omega_Q} \over {ea}},  \label{eq2.4}
\end{equation}
which is equal to 2.0$\times 10^5$V/cm for the aforementioned parameter 
values. 
Throughout the paper we assume the periodic boundary condition in order to 
avoid the end-point effects. 

The static polaron solution is determined by solving the self-consistent 
equation for the bond variable $y_n$ derived so as to minimize the 
total energy of the system in the absence of the external field; 
\begin{equation}
y_n = -{{2\alpha} \over K} \psi(n+1)\psi(n) + {{2\alpha} \over {NK}} \sum_{n'} 
\psi(n'+1)\psi(n'), \label{eq2.5}
\end{equation}
where $\psi$ is the electronic ground state wave function for the Hamiltonian 
$H_{\rm ele}$ (with $A=0$) which involves bond variables $\{ y_n \}$. 
The second term on the right hand side is introduced in order to certify the 
relation $\sum_n y_n = 0$ caused by the periodic boundary condition. 

The time evolution of the system in the presence of an external electric 
field is described by the time-dependent Schr\"odinger equation for the 
electronic wave function and the equations of motion for the lattice 
displacements $\{ u_n \}$. The former is written as 
\begin{eqnarray}
{\rm i}\hbar {{\partial} \over {\partial t}}\psi(n,t) & = & -(t_0-\alpha y_n)
{\rm e}^{{\rm i}\gamma A}\psi(n+1,t) \nonumber \\
 & & - (t_0-\alpha y_{n-1}){\rm e}^{-{\rm i}\gamma A}\psi(n-1,t), \label{eq2.6}
\end{eqnarray}
while the latter is expressed as
\begin{equation}
M{{{\rm d}^2u_n} \over {{\rm d}t^2}} = -{ \partial \over {\partial u_n}}
\left\{ \varepsilon_{\rm ele}(\{ y_n \}) + {K \over 2}\sum_{n'} y_{n'}^2 
\right\}, \label{eq2.7}
\end{equation}
where $\varepsilon_{\rm ele}(\{ y_n \})$ represents the electronic energy 
which is treated as a Born-Oppenheimer type adiabatic potential for the 
lattice motion. 

The time dependent Schr\"odinger equation is solved by 
using the fractal path integral method.~\cite{rf:15,rf:16} 
The Hamiltonian matrix is divided into two parts, one of which involves 
only the transfer processes between $(2n-1)$-th and $2n$-th sites with 
$n$ an arbitrary natural number and the other only those between 
$2n$-th and $(2n+1)$-th sites. As the fractal path integral formula, 
we adopt the simplest one which is correct up to the second order of 
the time mesh $\Delta t$ for the numerical simulation, which is chosen to be 
sufficiently small compared to the characteristic time of the lattice 
vibration, $\omega_Q^{-1}$. 
The electric field is slowly switched on and off with a certain duration 
time $\tau$ in order to avoid disturbances due to sudden 
switching.~\cite{rf:5,rf:17} The details of the time dependence of the 
field are described in ref.~\cite{rf:5}.

\section{Results of Numerical Simulations}

The spatial structure of the static acoustic polaron is shown in 
Fig.~\ref{Fig:1}, 
where the total number of the lattice sites is $N = 200$; the top is the 
electron density $\rho_n(=|\psi(n)|^2)$, the middle the lattice 
displacement $u_n$, the bottom the bond variable $y_n$ and the abscissa of 
all the figures being the site number $n$. The center of the polaron is 
chosen to be at $n=100$. Furthermore the center of mass of the lattice 
system is fixed at the same point; this is allowed because of the 
translational invariance of the SSH Hamiltonian.  Note that the bond 
variable $y_n$ takes a finite 
(non-vanishing) value even in the region far from the polaron center. 
This is due to the finiteness of the system size and the periodic boundary 
condition.  This is reflected also in the behavior of the lattice displacement 
$u_n$ which has a finite slope in the region far from the polaron. 
Apart from this finite size effect, the behaviors of $u_n$ and $y_n$ are 
in good agreement with the analytic result by Wilson~\cite{rf:2} in the 
continuum limit. The half width of the polaron is about 20 times the lattice 
constant. The charge density is well localized within the polaron width. 
\begin{figure}

\caption{The spatial structure of an acoustic polaron obtained in a 
finite chain of the SSH model with periodic boundary condition. The electron 
density $\rho_n$ (the top), the lattice displacement $u_n$ (the middle) 
and the bond variable $y_n$ (the bottom) are plotted as functions of 
the site number $n$. The total number of sites is $N=200$ and the values 
of parameters are given above eq.~(2.4). $u_n$ and $y_n$ are both 
scaled by the lattice constant $a$.} 

\label{Fig:1}

\end{figure}

In the following we show mainly the results of numerical simulations where the 
acoustic polaron shown in Fig.~\ref{Fig:1} is accelerated by an electric field 
with strength 0.005$E_0$. The time mesh $\Delta t$ is chosen to be 
0.0025$\omega_Q^{-1}$ and the duration time of switching-on is set to be 
200$\omega_Q^{-1}$. 

\subsection{Saturation of velocity}

The velocity of the polaron is calculated from the time 
dependence of the center of mass, which can be obtained by regarding the 
electron density or the bond variable\cite{com:1} as the probability weight 
for the position. Details on how to calculate the center of mass are 
described in ref.~\cite{rf:4}. The velocity of the polaron shows 
saturation if we continue to apply the electric field, similarly as 
in the case of the soliton in PA. An example is depicted in 
Fig.~\ref{Fig:2}, where the polaron velocity $v_{\rm p}$ scaled by the sound 
velocity $v_{\rm s}$ is given as a function of the time scaled by 
$\omega_Q^{-1}$; the upper graph is obtained from the electron density 
and the lower from the bond variable. The dashed line indicates the 
saturation velocity which is numerically estimated in the following way. 
\begin{figure}
\caption{The time dependence of the polaron velocity $v_{\rm p}$ 
obtained from the electron density (the upper graph) and from the bond 
variable (the lower graph). The system size is $N=200$ and the field 
strength is 0.005$E_0$, which is slowly switched on with a duration time 
200$\omega_Q^{-1}$. The horizontal dashed line indicates the saturation 
velocity. }
\label{Fig:2}
\end{figure}

In order to estimate the saturation velocity we fit the numerical 
data to the following form, 
\begin{equation}
v_{\rm p}(t) = v_{\rm sat} - v_0\exp(-t/T_{\rm sat}), \label{eq3.1}
\end{equation}
for the time interval $6\times 10^3 < t\omega_Q < 11\times 10^3$. The best 
fit curve is given by a thin line in Fig.~\ref{Fig:2}. The data for the small 
$t$ region are not used since they would not fit to eq.~(\ref{eq3.1}), 
and those for very large $t$ region are also not used because of 
disturbances which may be due to excitations of vibrational modes 
created by the motion of the polaron. From this fitting, we obtain 
$v_{\rm sat} = 0.90v_{\rm s}$, $T_{\rm sat} = 7.0 \times 10^3\omega_Q^{-1}$ 
for the upper graph of Fig.~\ref{Fig:2} and $v_{\rm sat} = 0.87v_{\rm s}$, 
$T_{\rm sat} = 4.0\times 10^3\omega_Q^{-1}$ for the lower graph. According to 
the analytic work by Wilson~\cite{rf:2} and the semi-phenomenological 
treatment by Kuwabara {\it et al.},~\cite{rf:18} the saturation velocity 
should be equal to $v_{\rm s}$. The present simulation has given a smaller 
value. The reason will be discussed later. We have carried out similar 
simulations for different values of the electric field strength (the same 
duration time) and confirmed that 
$v_{\rm sat}$ is almost independent of the strength of the applied field. 
On the other hand, the saturation time $T_{\rm sat}$ is found to depend 
linearly on the logarithm of the field strength, similarly as in 
the case of the soliton.~\cite{rf:4} 

\subsection{Width of polaron}

The width of the polaron can be estimated from the variance of the 
distribution function of the position which is given by the electron 
density or by the bond variable.~\cite{rf:5} We denote the polaron 
width obtained from the electron density by $\xi_{\rm ch}$ and that 
from the bond variable by $\xi_{\rm lat}$.~\cite{com:2}  
The results of simulations 
are summarized in Fig.~\ref{Fig:3} where the width scaled by its initial 
(static) value is given as a function of time. The relative changes of the 
widths obtained from the electron density and from the bond variable 
are almost the same in contrast to the case of the soliton in PA 
where the relative change of the width obtained from the excess charge 
distribution was about a half of that obtained from the bond order 
parameter.~\cite{rf:5}  Since we have calculated both the velocity and 
the width as functions of time, it is possible to eliminate the time 
and express the width as a function of velocity. The results are given in 
Fig.~\ref{Fig:4}. Both of $\xi_{\rm ch}$ and $\xi_{\rm lat}$ look to decrease 
to zero at the sound velocity, which is consistent with the analytical 
results.~\cite{rf:2,rf:18} 
\begin{figure}
\caption{The time dependence of the polaron width $\xi_{\rm ch}$ 
obtained from the electron density profile and $\xi_{\rm lat}$ obtained 
from the bond variable. Both are scaled by their static initial values. 
Data are taken from the simulation giving Fig.~2. Note that 
$\xi_{\rm lat}$ is not well-defined for large values of $t$ because of lattice 
vibrational modes excited due to the motion of the polaron. 
}
\label{Fig:3}
\end{figure}
\begin{figure}
\caption{The velocity dependence of the polaron width obtained by 
combining Fig.~2 and Fig.~3. The upper graph is 
obtained from the electron density, and the lower from the bond variable. 
}
\label{Fig:4}
\end{figure}

\subsection{Energy and effective mass}

The energy of the system shows also an anomalous behavior when the polaron 
velocity approaches the sound velocity. The electronic energy decreases 
with increasing polaron velocity. This is consistent with the behavior of 
the width, which is a decreasing function of the velocity as seen 
in Fig.~\ref{Fig:4}. 
The narrower the width, the deeper the effective potential felt by the 
electron and as a result the binding energy can be increased, which 
leads to the decrease of the electronic energy.  This fact indicates 
that the polaron becomes more stable when it moves, at least from the 
electronic point of view. In contrast to the electronic energy, the lattice 
energy (the sum of the lattice potential and kinetic energies) increases 
with increasing velocity due to the increase of the lattice kinetic energy 
and the increase of the lattice deformations. The total energy of the 
system looks to diverge when the velocity is increased towards the 
saturation velocity. 

Similarly as in the previous subsection, we can derive the relation 
between the energy and the velocity from the time dependences of the 
two quantities. The velocity dependences of the electronic energy 
$\varepsilon_{\rm ele}$, the lattice potential energy 
$\varepsilon_{\rm lp}$, the lattice kinetic energy $\varepsilon_{\rm lk}$ 
and the total energy $\varepsilon_{\rm tot}$ are shown in Fig.~\ref{Fig:5}, 
where the difference from the static values are given as functions of the 
velocity calculated from the electron density. The lattice potential and 
kinetic 
energies correspond to the first and the second terms of eq.~(\ref{eq2.3}), 
respectively. According to the analytic works, $\varepsilon_{\rm lp}$ and 
$\varepsilon_{\rm lk}$ diverge as $(v_{\rm s}-v_{\rm p})^{-3}$ when 
$v_{\rm p}$ approaches $v_{\rm s}$, while $|\varepsilon_{\rm ele}|$ 
diverges as $(v_{\rm s}-v_{\rm p})^{-2}$. Therefore the total energy 
behaves as $(v_{\rm s}-v_{\rm p})^{-3}$ when $v_{\rm p}$ tends to 
$v_{\rm s}$. The overall behaviors of energies as functions of velocity 
are quite well described by these analytic predictions as discussed 
in ref.~\cite{rf:18}. 
\begin{figure}
\caption{The velocity dependences of various energies obtained 
from the simulation giving Fig.~2 and Fig.~3. 
The electronic energy $\varepsilon_{\rm ele}$, the lattice potential 
energy $\varepsilon_{\rm lp}$, the lattice kinetic energy 
$\varepsilon_{\rm lk}$ and the total energy 
$\varepsilon_{\rm tot}$ are plotted as functions of the polaron velocity 
$v_{\rm p}$ determined from the electron density; only the differences 
from the static values are given. Inset is 
$\varepsilon_{\rm tot}$ in the small $v_{\rm p}$ region, the dashed curve 
being the best fit in the region $v_{\rm p}/v_{\rm s} < 0.5$ to the sixth 
order polynomial; the coefficient of the quadratic term is proportional 
to the effective mass of the polaron. 
}
\label{Fig:5}
\end{figure}

We can also estimate the effective mass of the polaron from the velocity 
dependence of the total energy. As shown in the inset of Fig.~\ref{Fig:5}, we 
have made a fitting of the data for $v_{\rm p}/v_{\rm s} < 0.5$ to the form, 
\begin{equation}
\varepsilon_{\rm tot} = {{m_{\rm p}v_{\rm s}^2} \over 2}\left( {{v_{\rm p}} 
\over {v_{\rm s}}}\right)^2 + C_4 \left( {{v_{\rm p}} \over 
{v_{\rm s}}}\right)^4 + C_6 \left( {{v_{\rm p}} \over {v_{\rm s}}}\right)^6 . 
\label{eq3.2}
\end{equation}
In this expression, $m_{\rm p}$ has a meaning as the effective mass of the 
polaron. From the least square fitting we obtain 
$m_{\rm p} = 1.0\times 10^2m_{\rm e}$ where $m_{\rm e}$ is the bare electron 
mass. This value agrees quite well with what is expected from the analytic 
calculations.~\cite{rf:2,rf:18} This rather heavy mass should be compared 
with the rather light effective mass of the soliton in PA 
which is $3 \sim 4 m_{\rm e}$.~\cite{rf:3,rf:6} 
The main reason of this difference will be due to the large difference 
in the mass of an ion unit for one lattice site, although the effective 
mass of the localized nonlinear excitation such as the soliton in 
PA or the acoustic polaron treated here is not directly 
related to the mass of the ion unit.  It would be worthwhile to note that 
the value of the transfer integral $t_0$ used in the present calculation 
is much smaller than that used in the similar calculations in PA; the latter 
is about four times larger than the former. This may be one of the reasons of 
the large difference in the effective masses of the present acoustic polaron 
in PDA and the soliton in PA.

\section{Linear Mode Analysis} 

Linear modes around an acoustic polaron can be obtained similarly 
as those around a soliton or a polaron in 
PA.~\cite{rf:8,rf:9,rf:11,rf:12} The explicit formulation 
for the linear mode analysis in the case of the SSH model has been 
precisely described in ref.~\cite{rf:12}.  Therefore we do not 
state it again here and express only the results for the case 
with $N=200$. 
In the following, the linear modes represent small deviations of the 
lattice displacements $\{ u_n\} $ from their static values in the presence 
of an acoustic polaron. 

In the linear mode analysis, we obtain two degenerate zero-frequency modes; 
one is a uniform mode which exists in any case because of the translational 
invariance of the original Hamiltonian and the other is a so-called 
Goldstone mode which describes the translational motion of the acoustic 
polaron.  The latter is shown in Fig.~\ref{Fig:6}. It has an essentially 
localized 
character, although it has a finite flat value in the region far from 
the polaron because of the finite size effect discussed in the previous 
section. In fact we have confirmed that the finite value outside the 
polaron region decreases proportionally to the inverse of $N$ by carrying 
out similar analyses for different system sizes. The fact that the 
eigenfrequency of this Goldstone mode is zero means that there is no pinning 
effect due to the discreteness of the lattice. This may be related to the 
very high mobility observed for the excess charges created by photo-excitation 
of PDA. 
\begin{figure}
\caption{The Goldstone mode obtained in the linear mode analysis. This 
mode corresponds to the translational motion of the polaron. 
}
\label{Fig:6}
\end{figure}

It is worthwhile to note 
that we could not find any other localized mode. This is in contrast 
to the case of the soliton in PA, where there are two more 
localized modes with optical character other than the Goldstone mode 
corresponding the translational motion of the soliton. 
In order to confirm that there is only one localized mode in the case of 
the acoustic polaron, we have performed the phase shift analysis of the 
extended modes. Although no direct analogy with the quantum mechanical 
potential problem is expected in this case, we can get a certain information 
about the localized mode from the wave number dependence of the phase shift, 
since it is related to the modification of the density of modes~\cite{rf:19} 
due to the presence of the polaron. 

\subsection{Phase shift analysis}

Extended modes are expressed in the form of plane waves in the region 
far from the polaron center, since the linear modes of the lattice vibration 
is affected by the presence of the polaron only in the region near to it. 
The extended modes with even and odd parity approach asymptotically to 
the following forms, 
\begin{eqnarray}
g_{\rm e}(n) & = & A_{\rm e} \cos \left[ k\left( n-{N \over 2}\right) a+
{{\delta_{\rm e}} \over 2} \right] , \label{eq4.1} \\
g_{\rm o}(n) & = & A_{\rm o} \sin \left[ k\left( n-{N \over 2}\right) a+
{{\delta_{\rm o}} \over 2} \right] , \label{eq4.2}
\end{eqnarray}
on the right hand side of the polaron, where we have chosen the center of the 
polaron at $n=N/2$.  Here $\delta_{\rm e}$ and 
$\delta_{\rm o}$ are the phase shifts of the extended modes with even 
and odd parity, respectively. The phase shifts and the wave numbers 
are calculated through the least square fitting to the above-mentioned 
function with three parameters in the region $3N/4 \le n \le N $. In this way 
we can derive a relation between the phase shift and the wave number. 
The results are summarized in Fig.~\ref{Fig:7}. 
\begin{figure}
\caption{The wave number dependence of the phase shift of the extended 
modes with even parity $\delta_{\rm e}$ and odd parity $\delta_{\rm o}$. 
}
\label{Fig:7}
\end{figure}

The phase shift for the even parity modes is linearly approaching to 
$2\pi$ for $k$ tending to zero, while that for the odd parity modes has 
a maximum as a function of $k$ and decreases to zero as $k$ goes to zero. 
Levinson's theorem for one-dimensional Schr\"odinger equation tells us that 
the zero limit value of the phase shift is related to the number of 
localized state as $\delta_{\rm e}(0) = (2n_{\rm e}-1)\pi$ and 
$\delta_{\rm o}(0) = 2n_{\rm o}\pi$ for even and odd parity, respectively, 
where $n_{\rm e}$ and $n_{\rm o}$ represent the number of localized states 
with even and odd parity, respectively.~\cite{rf:14} 
The present result for $\delta_{\rm e}$ does 
not coincide with this. However this is not astonishing, since in the 
present problem the eigenvalue (=the square of the eigenfrequency) 
becomes never negative in contrast to the negative eigenenergies of 
localized states in the Schr\"odinger equation and since the 
localized mode in the present problem is strictly localized only in 
the infinite system and has a character as an extended mode in finite 
systems.  In the linear mode analysis in the case of the soliton or 
the polaron in PA, the existence of the analogy with the 
Schr\"odinger equation was pointed out.~\cite{rf:8,rf:9,rf:11,rf:12} 
The situation there is completely different from the present one, since 
the reference frequency in that case is the renormalized optical 
phonon frequency and the frequencies of the localized modes are 
smaller than this reference frequency. 
If we regard the Goldstone mode shown in Fig.~\ref{Fig:6} as an extended 
mode the behavior of $\delta_{\rm e}$ near $k=0$ is quite reasonable, 
because this mode can be regarded as a uniform extended mode with a 
phase shift $2\pi$ as is easily seen from Fig.~\ref{Fig:6}. The behavior of 
$\delta_{\rm o}$ is consistent with the fact that there is no localized 
mode with odd parity. It should be noted that phase shifts for even and 
odd parity coincide with each other for larger values of $k$. This means 
that the lattice vibrations with larger $k$ suffer reflectionless scatterings 
from the polaron.~\cite{rf:20} This will be important when we consider the 
interactions between phonons and the acoustic polaron. 

We have performed similar phase shift analyses for different values of 
the coupling constant $\alpha$ in order to check whether the above conclusion 
can be affected by the change of the coupling constant. The obtained results 
are qualitatively the same as Fig.~\ref{Fig:7}. In general the stronger 
coupling pushes up the phase shifts for larger $k$ but does not change the 
behaviors near $k=0$. The peak position of $\delta_{\rm o}$ and the 
reflectionless region are moved to larger $k$ as the coupling constant 
increases. 

\subsection{Amplitude mode} 

As discussed in the previous subsection we could not find any localized 
mode with odd parity. This means that there is no amplitude mode which has 
been confirmed to exist in the case of the soliton or polaron in PA. 
The amplitude mode would determine the oscillation of the width of the 
acoustic polaron if any. The lack of the amplitude mode makes us expect 
that the moving polaron may not show any oscillation of its width. 
In order to see whether this is the case, we have carried out 
a simulation where the electric field is switched off at a certain 
time.~\cite{rf:7} 

In Fig.~\ref{Fig:8}, the result for $\xi_{\rm lat}$ is depicted. An electric 
field with a strength $E=0.002E_0$ is switched on with a duration time 
$\tau = 125\omega_Q^{-1}$ and switched off with the same duration time 
just after it reaches the stationary value; namely the field is 
completely switched off at $t=250\omega_Q^{-1}$. Although the 
amplitude is not very large, we see an almost periodic oscillation of the 
width. This can be understood if we look at the lowest extended mode with 
odd parity shown in Fig.~\ref{Fig:9}. Its shape is topologically the same as 
what is expected for the amplitude mode (the dash-dotted line in 
Fig.~\ref{Fig:9}). Therefore 
if this mode is excited the polaron width will oscillate with time. This is 
confirmed by the following facts. First of all, the period $T_{\rm dyn}$ 
of the width oscillation observed in Fig.~\ref{Fig:8} is equal to 
390$\omega_Q^{-1}$, 
while the period of the mode depicted in Fig.~\ref{Fig:9} is 
$T_{\rm lin}=400\omega_Q^{-1}$ which is obtained from the eigenfrequency 
calculated in the linear mode analysis. Secondly, in order to confirm that 
this agreement is not accidental, we performed similar calculations for 
different system sizes. The results are $T_{\rm lin} = 210\omega_Q^{-1}$ 
and $T_{\rm dyn} = 200\omega_Q^{-1}$ for $N=100$, and 
$T_{\rm lin} = 800\omega_Q^{-1}$ and $T_{\rm dyn} = 840\omega_Q^{-1}$ for 
$N=400$. The agreement between $T_{\rm lin}$ and $T_{\rm dyn}$ seems 
satisfactory. We can therefore conclude that the oscillation of the width will 
die out in the thermodynamic limit. If the effective chain length becomes 
finite e.g. by the presence of defects, however, it may be possible to 
observe the width oscillation. 
\begin{figure}
\caption{The behavior of the polaron width $\xi_{\rm lat}$ (obtained 
from the bond variable) after switching off the 
electric field.  The vertical broken line indicate the time at which the 
field vanishes completely. The strength of the electric field is 0.002$E_0$ 
and the field is switched on and off slowly with a duration time 
125$\omega_Q^{-1}$; as a result the field vanishes at $t=250\omega_Q^{-1}$. 
}
\label{Fig:8}
\end{figure}
\begin{figure}
\caption{The lowest extended mode with odd parity (solid curve). 
The dash-dotted curve shows what would be expected if there were a localized 
amplitude mode.}
\label{Fig:9}
\end{figure}

\section{Summary and Discussion}

In this paper we reconsidered the dynamics of an acoustic polaron in 
PDA, which was analytically treated by Wilson.~\cite{rf:2}
We carried out numerical simulations by using the SSH model and applying 
formulations developed for studying the soliton or polaron in PA. 
The main difference between the two problems lies in the number of electrons: 
When we consider the acoustic polaron, we assume there is only one electron in 
the conduction band, whereas the conduction band is half-filled by electrons 
in PA. 

The saturation of the polaron velocity at the value near to the sound 
velocity has been confirmed in the simulation where the external electric 
field is kept to be applied. The reduction of the polaron width with 
increasing velocity has been also confirmed; the result of 
simulation indicates that the width will vanish at the sound velocity. 
The behavior of energy suggests a divergence at the saturation velocity, 
which is consistent with the analytic studies.~\cite{rf:2,rf:19} 
Although the saturation velocity estimated from the time dependence of the 
velocity is slightly smaller than the sound velocity, the behaviors of 
the width and the energy as function of the velocity clearly indicate that 
the saturation velocity is exactly the sound velocity. 
The fact that the saturation velocity obtained in the dynamical simulation 
is smaller than the sound velocity is partly due to the friction created 
by the interaction between the polaron and the lattice vibrations excited 
by the motion of the polaron, and partly due to that it takes very long time 
to reach the saturation velocity; it may be too long to be confirmed this 
type of simulations. 

Furthermore in order to understand the dynamics of the acoustic polaron 
more deeply, we have performed the linear mode analysis. As a result we 
have found only one localized mode corresponding to the translational motion 
of the polaron. The phase shift analysis of the extended modes is found to 
be consistent with this. Although an exactly same relation between 
the zero-limit value of the phase shift and the number of localized modes as 
given in Levinson's theorem for the one-dimensional Schr\"odinger equation 
was not found in the present phase shift analysis, it has been argued that 
this is not astonishing because of the difference between the two situations. 

The acoustic polaron was introduced by Wilson~\cite{rf:2} to explain the 
experimentally observed behaviors of the excess charge generated in 
PDA by means of photo-excitation.~\cite{rf:1} These carriers 
have an ultra-high mobility and their velocity is constant (less than the 
sound velocity) in low electric field. These characteristics are essentially 
understood by the behavior of the acoustic polaron. In the experiment, more 
than one electrons may be excited per chain. In this case we have to consider 
bipolarons or more sophisticated objects, for which the mutual interactions 
between electrons cannot be neglected. It will be interesting to study the 
cases with two or more electrons in the conduction band by a similar method 
used in the present work. This is left for the future work along with the 
study of the electron-electron interaction effects.

\section*{Acknowledgements}

The authors wish to express their gratitude to Professor T. Ohtsuki and 
M. Kinoshita for fruitful discussions. This work was partially financed by 
Grant-in-Aid for Scientific Research from the Ministry of Education, 
Science and Culture, No. 05640446.



\begin{thebibliography}{99}
\bibitem{com:1} In this case, we shift the bond variable so that the flat part 
far away from the polaron position may become zero.
\bibitem{com:2} In fact the estimation of $\xi_{\rm lat}$ by the variance is 
not very reliable for long time lapse, because of disturbances in the bond 
variables appearing in the regions far from the polaron center due to the 
motion of the polaron.  We have, therefore, estimated $\xi_{\rm lat}$ by 
fitting the bond variable $y_n(t)$ to 
$$y_n(t) = y_0 {\rm sech}^{-2}[(n-n_0)a/\xi_{\rm lat}] + c $$
with $y_0$, $\xi_{\rm lat}$, $n_0$ and $c$ as fitting parameters. It has been 
confirmed that the two estimations are consistent for short time lapse.
\end{thebibliography}
\end{document}